# MULTIMODAL INFORMATION FUSION FOR THE DIAGNOSIS OF DIABETIC RETINOPATHY

Li Yihao[1, 2]; Al Hajj Hassan[1, 2]; Conze Pierre-Henri[3, 2]; El Habib Daho Mostafa[1, 2]; Bonnin Sophie[4]; Ren Hugang[5]; Manivannan Niranchana[5]; Magazzeni Stephanie[5]; Tadayoni Ramin[6]; Lamard Mathieu[1, 2]; Quellec Gwenole[2]

1. Universite de Bretagne Occidentale, Brest, Bretagne, France. 2. INSERM, LaTIM, UMR 1101, France.
3. IMT Atlantique, Brest, France. 4. Hopital Rothschild, Paris, Île-de-France, France.
5. Carl Zeiss Meditec Inc, Dublin, CA, United States. 6. Hopital Lariboisiere, Paris, Île-de-France, France.

## Context

Diabetes is a chronic disease characterized by excess sugar in the blood and affects 422 million people worldwide, including 3.3 million in France. One of the frequent complications of diabetes is diabetic retinopathy (DR): it is the leading cause of blindness in the working population of developed countries. As a result, ophthalmology is on the verge of a revolution in screening, diagnosing, and managing of pathologies. This upheaval is led by the arrival of technologies based on artificial intelligence.

The "Evaluation intelligente de la rétinopathie diabétique" (EviRed) project uses artificial intelligence to answer a medical need: replacing the current classification of diabetic retinopathy which is mainly based on outdated fundus photography and providing an insufficient prediction precision. EviRed exploits modern fundus imaging devices and artificial intelligence to properly integrate the vast amount of data they provide with other available medical data of the patient. The goal is to improve diagnosis and prediction and help ophthalmologists to make better decisions during diabetic retinopathy follow-up.

In this study, we investigate the fusion of different modalities acquired simultaneously with a PLEX® Elite 9000 (Carl Zeiss Meditec Inc. Dublin, California, USA), namely 3-D structural optical coherence tomography (OCT), 3-D OCT angiography (OCTA) and 2-D Line Scanning Ophthalmoscope (LSO), for the automatic detection of proliferative DR.

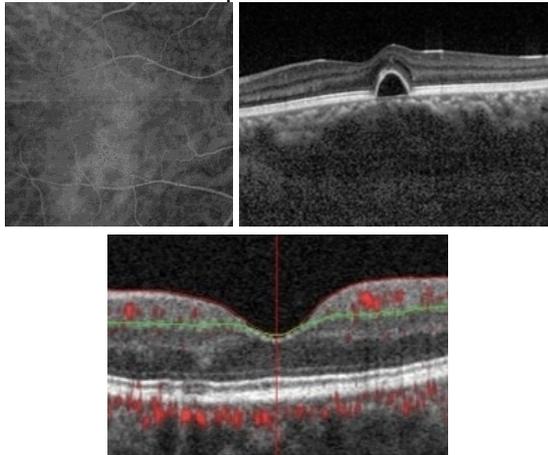

Figure 1: Three modalities: LSO (top left), structural OCT (top right), OCT-A (bottom – red indicates presence of flow).

## Methods

In this study, 151 OCT volumes from 64 diabetic patients were collected. This collection was divided as follows: 88 acquisitions (from 31 patients) for training, 28 acquisitions (from 14 patients) for validation and 35 acquisitions (from 19 patients) for testing. DR severity level, according to the International Clinical Diabetic Retinopathy Disease Severity Scale (ICDR) scale, was graded by a retina specialist using fundus photographs: 30 acquisitions (including 16 in the train set, 5 in the validation set and 9 in the test set) had proliferative DR. Three fusion methods were evaluated: early fusion, intermediate fusion, and hierarchical fusion.

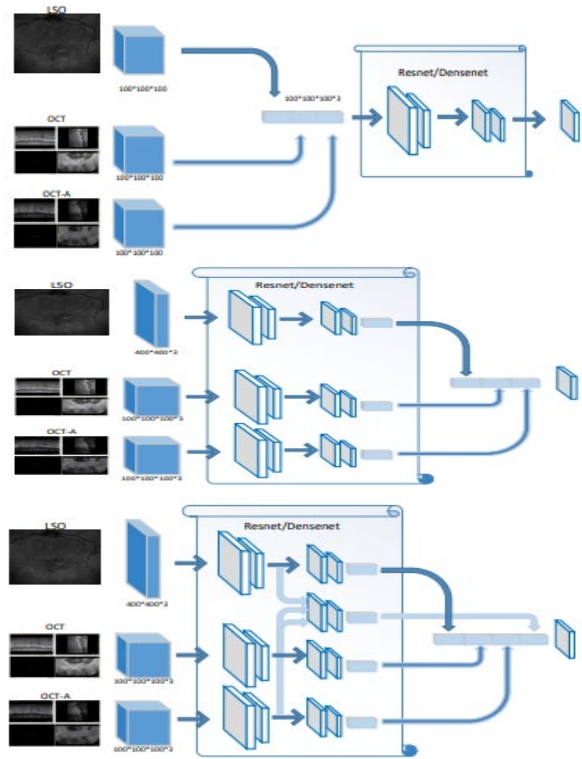

Figure 2: Three fusion methods: early fusion (top), intermediate fusion (middle) and hierarchical fusion (bottom).

The network structure of the different fusion methods is shown in Figure 2. In the early fusion scheme, multi-modality images are fused channel by channel as multi-channel inputs to learn a fused feature representation, and then train the segmentation network. In the intermediate fusion scenario, each modality image is used as a single input of a single classification network. The outputs of the individual networks are then integrated to get the final result [1]. In the hierarchical fusion strategy, each modality image is used as a single input of a single classification network, and then these learned individual feature representations are fused in the deeper layers of the network. Finally, the fused result is fed to the decision layer to obtain the final result [2].

| Method | Backbone | AUC | Sensitivity | Specificity | Improvement |
| --- | --- | --- | --- | --- | --- |
| **Single modality (Structure)** | Resnet101 | 0.859 | 0.78 | 0.77 | Baseline |
| **Single modality (Flow)** | Densenet169 | 0.816 | 0.78 | 0.85 | -0.043 |
| **Single modality (LSO)** | Densenet121 | 0.662 | 0.67 | 0.74 | -0.197 |
| **Hierarchical fusion** | Densenet121 | **0.911** | 0.86 | 0.88 | +0.052 |
| **Early Fusion** | Densenet121 | 0.865 | 0.78 | 0.85 | +0.006 |
| **Intermediate Fusion** | Densenet121 | 0.744 | 0.67 | 0.85 | -0.115 |

*Table 1: Results of different fusion methods.*

The following backbones were investigated for each method: Resnet50, Resnet101, Densenet121, and Densenet169. The Area under the ROC Curve (AUC) was used to assess classification performance. These fusion methods were compared to the classification of a single modality separately.

## Results

Using a single modality, the structure data achieved the best performance: AUC reaches 0.859 using Resnet101; this is our baseline. The Flow data reached an AUC of 0.816, using Densenet169. The LSO data reached an AUC of 0.662, using Densenet121. Hierarchical fusion achieves the best results, AUC reaches 0.911 using Densenet121, with an increase in AUC of over 0.052 compared to baseline. While the AUC increase for Early Fusion, on Densenet121, was around 0.006. Intermediate Fusion on Densenet121 performed worse than baseline.

## Discussion

The results show the effectiveness of our multimodal fusion method for proliferative DR detection. The hierarchical fusion method is currently showing promising results. In particular, it outperforms the detection in a single modality. However, these experiments will have to be replicated in a larger dataset to achieve clinically useful detection performance.

## Acknowledgements

The work takes place in the framework of the ANR RHU project Evired. This work benefits from State aid managed by the French National Research Agency under "Investissement d'Avenir" program bearing the reference ANR-18-RHUS-0008.